\documentclass{article}
\usepackage{amsmath}

\input{tcilatex}
\begin{document}

\title{Transition from Solitons to Solitary Waves in the Fermi-Pasta-Ulam
Lattice}
\author{Zhenying Wen$^{\ast }$, Jun Tao, Nian Wei \\
Center of Theoretical Physics, College of Physical Science and\\
Technology, Sichuan University, Chengdu 610065, China}
\maketitle

\begin{abstract}
In this paper, we study the smooth transition from solitons to solitary
waves in localization, relation between energy and velocity, propagation and
scattering property in the Fermi-Pasta-Ulam lattice analytically and
numerically. A soliton is a very stable solitary wave that retains its
permanent structure after interacting with other solitary waves. A soliton
exists when the energy is small, and it becomes a solitary wave when the
energy increases to the threshold. The transition could help to understand
the distinctly different heat conduction behaviors of the Fermi-Pasta-Ulam
lattice at low and high temperature.

\end{abstract}

\bigskip \textbf{I. INTRODUCTION}

The history of solitary waves can be traced back to 1834, when it was first
discovered by Russell on the Union Canal in Scotland \cite{1}. Several
decades later Korteweg--de Vries equation was provided which explained the
phenomenon mathematically \cite{2}. More than a century later, Fermi, Pasta
and Ulam (FPU) investigated the fundamental problem of energy equipartition
and ergodicity in nonlinear systems and discovered FPU recurrence \cite{3},
which has led eventually to the soliton concept \cite{4}. Since then
solitary waves have been studied in various different fields of physics like
solid-state physics, quantum theory, nonlinear optics, fluid dynamics,
biophysics, etc. \cite{5,6}.

Distinguishing between solitons and solitary waves is sometimes of
importance such as studies on energy transport. Solitons are localized
solutions of integrable equations, and they are very special types of
solitary waves which do not alter form and only have a phase shift after
collision with other solitons. Solitary waves are associated with
nonitegrable equations, and they are deformed and energy exchange may occur
when solitary waves interact with one another \cite{7}. In the long
wavelength approximation, the FPU-$\alpha $ and FPU-$\beta $ lattices lead
to the KdV and modified KdV equations, respectively, in which the exact
soliton solutions exist. The stability of soliton in the FPU lattice at low
energy is demonstrated \cite{8,9,10,11}. At high energy level, it becomes
solitary wave, which makes attribution to heat conduction \cite{12,13}
because of the inelastic scattering between each other. Up to now, very
little work has been done on the study of changing from solitons to solitary
waves in the FPU lattice.

In this work, we focus on the transition from solitons to solitary waves in
the FPU lattice. Previously in numerical work solitons are constructed by
using analytical soliton solution directly \cite{14} or starting with a
special initial condition \cite{4}. Here we obtain solitons in the FPU
lattice for the first time with momentum excitation, which is the widely
used method to excite solitary waves \cite{12,13,15,16,17}. Soliton
transforms to solitary wave when energy increases larger than the threshold.
Solitons and solitary waves are different in relation between energy and
velocity, propagation and scattering behavior. Solitons and solitary waves
maintain shape and energy when travelling; kinetic and potential energy
remain unchanged for solitons while they are varying periodically for
solitary waves. The scattering of solitons is elastic. When two solitary
waves collide with each other, both of them lose energy or they exchange
energy, and extra wave packets are excited. The solitary wave properties
become similar to those in the pure anharmonic lattice when energy tends to
infinity. Finally the different heat conduction behaviors of the FPU lattice
at low and high temperature are discussed with the concept of solitons and
solitary waves.

\bigskip \textbf{II. ANALYSIS\ IN THE\ WEAKLY\ NONLINEAR\ LIMIT}

We consider the Hamiltonian for the FPU lattice

\begin{equation}
H=\sum\limits_{n}H_{n},H_{n}=\frac{p_{n}^{2}}{2}+V(q_{n+1}-q_{n})  \label{1}
\end{equation}

\begin{equation}
V(q)=\alpha q^{2}/2+\beta q^{4}/4  \label{2}
\end{equation}%
\ \ \ \ \ \ \ \ where $p_{n}$ denotes the momentum and $q_{n}$ denotes the
displacement from equilibrium position for the $nth$ atom. Dimensionless
units are used, such that the masses, the linear and nonlinear force
constants, and the lattice constant are all chosen to be unity. The
parameters $\alpha $ and $\beta $ are the harmonic and anharmonic force
constant, respectively. Here $\alpha =1$ and $\beta =1$ for the FPU lattice
throughout the paper.

The equations of motion of the FPU system are

\begin{equation}
\ddot{q}%
_{n}=(q_{n+1}-q_{n})+(q_{n-1}-q_{n})+(q_{n+1}-q_{n})^{3}+(q_{n-1}-q_{n})^{3}
\label{3}
\end{equation}%
In the long wavelength limit, the continuum approximation should be
applicable. The displacement $q_{n\pm 1}$ of the $\left( n\pm 1\right) th$
lattice is expanded as

\begin{equation}
q_{n\pm 1}=q\pm q_{x}+\frac{1}{2}q_{xx}\pm \frac{1}{6}q_{xxx}+\frac{1}{24}%
q_{xxxx}+\cdots  \label{4}
\end{equation}%
where $q\left( x,t\right) =q_{n}\left( t\right) $, and $x=n$. Substituting
Eq. (4) into Eq. (3) and neglecting higher order terms, we obtain the
continuous counterparts of Eq. (3):

\begin{equation}
q_{tt}=\left( 1+3q_{x}^{2}\right) q_{xx}+\frac{1}{12}q_{xxxx}  \label{5}
\end{equation}
\ \ 

\ We shall be interested in right-going waves, and introduce the new slow
variables

\begin{equation}
\xi =2\varepsilon \left( x-t\right) ,\tau =\varepsilon ^{3}t  \label{6}
\end{equation}%
and define

\begin{equation}
q\left( x,t\right) =\varphi \left( \xi ,\tau \right)  \label{7}
\end{equation}%
where $\varepsilon $ is a formal small parameter. Substituting Eq. (6) and
(7) into Eq. (5) and keeping terms in the order of $\varepsilon ^{4}$, we
obtain the following evolution equation for the function $\psi =\partial
\varphi /\partial \xi $:

\begin{equation}
\psi _{\tau }+\frac{3}{2}\psi ^{2}\psi _{\xi }+\frac{1}{24}\psi _{\xi \xi
\xi }=0  \label{8}
\end{equation}%
which is the well known modified Korteweg-de Vries equation. Eq. (8) is
derived from the FPU chain \cite{18,19} and it results in the soliton
solutions \cite{14,20,21}:

\begin{equation}
q_{n}=\varphi =-\sqrt{2/3}\arctan \{e^{A\sqrt{6}\left[ n-t-\left( A/2\right)
^{2}t\right] }\}+c  \label{9}
\end{equation}

\begin{equation}
p_{n}=\dot{q}_{n}=A\left[ 1+\left( A/2\right) ^{2}\right] \sec h\text{ }\{A%
\sqrt{6}\left[ n-t-\left( A/2\right) ^{2}t\right] \}  \label{10}
\end{equation}%
where $c$ is constant. Eq. (9) and (10) are valid if the higher derivatives
are neglected. It can be achieved if the following is fulfilled.

\begin{equation}
6A^{2}\ll 1  \label{11}
\end{equation}

The solutions for $q_{n}$ and $p_{n}$ are kink and bell shaped soliton
respectively and the soliton velocity is achieved:

\begin{equation}
v=1+(A/2)^{2}  \label{12}
\end{equation}%
The soliton energy is

\begin{equation}
E=\sum_{n}\frac{p_{n}^{2}}{2}%
+\sum_{n}[(q_{n+1}-q_{n})^{2}/2+(q_{n+1}-q_{n})^{4}/4]  \label{13}
\end{equation}
The first term of Eq. (13) denotes kinetic energy of soliton and the second
term is potential energy. In the long wavelength limit, potential energy is
very close to kinetic energy for soliton at low energy, so soliton energy
can be expressed as follow:

\begin{equation}
E=2\sum_{n}\frac{p_{n}^{2}}{2}  \label{14}
\end{equation}%
Since soliton energy is unchanged as soliton propagates, substituting Eq.
(10) at the time $t=0$ we obtain

\begin{equation}
E=\sum_{n}\{A\left[ 1+\left( A/2\right) ^{2}\right] \sec h\left( A\sqrt{6}%
n\right) \}^{2}  \label{15}
\end{equation}%
Replacing the sum by an integral in Eq. (15) due to continuum approximation
and using Eq. (11), we finally arrive to

\begin{equation}
E=\frac{2}{\sqrt{6}}A  \label{16}
\end{equation}%
Then we can get the relation between soliton energy and velocity from Eq.
(12) and (16) in the case of soliton energy $E\ll \frac{1}{3}$:

\begin{equation}
v=1+\frac{3}{8}E^{2}  \label{17}
\end{equation}%
So soliton velocity is close to acoustic velocity due to the small energy.

\textbf{III.\bigskip\ HARMONIC LIMIT AND ANHARMONIC LIMIT}

We obtain the harmonic lattice in the absence of the quartic term, i.e., $%
\alpha =1$ and $\beta =0$ for Eq. (1) and (2). The system is integrable and
can be analytically solved. Here acoustic velocity is 1.

It is the pure anharmonic lattice with $\alpha =0$ and $\beta =1$. In this
case the excitation, propagation and interaction of solitary waves have been
studied in Refs. 13,15, 22, 23. The system has an excellent scaling
property, from which we can obtain the relation between energy $E$ and
velocity $v$ of a solitary wave:%
\begin{equation}
v=aE^{1/4}  \label{18}
\end{equation}%
$a=0.68198$ from numerical simulations \cite{13}.

\bigskip \textbf{IV. TRANSITION FROM SOLITONS TO SOLITARY WAVES IN THE FPU
LATTICE}

Solitary waves can be excited by momentum kicks on the lattice. We apply a
kick $p_{1}$ on the first lattice at $t=0$ for static system\ under the free
boundary condition. In Fig. 1 we plot $p_{i}$ versus $i$ after a short time (%
$t=500$) for three kinds of lattices: the harmonic chain at $p_{1}=0.7$ (a),
the pure anharmonic chain at $p_{1}=0.7$ (b), the FPU chain at $p_{1}=1.5$
(c) and the FPU chain at $p_{1}=0.7$ (d). In the harmonic chain the
amplitude of the wave profile decreases while the width increases with time
continuously. In the pure anharmonic chain a solitary wave is excited
accompanied by the tail, which is made of several small solitary waves shown
in the inset of Fig. 1 (b) \cite{13}. In the FPU chain, at first the wave
front is connect with the other low amplitude excitations. If the imparted
kick is large enough, after a certain short time, the wave front separates
form the tail, because it moves faster. It propagates forward and becomes a
solitary wave, just like that in Fig. 1 (c). If the kick is small, the
excited wave packet looks like that in the harmonic chain in a short time
(Fig. 1 (d)). But actually It will show the essential difference after a
long time. Consider the lattice which is so long that the wave packet won't
reach the other end of the lattice in the long observing time. As the first
wave pulse travels, its width becomes wider and its peak becomes lower, as
shown in Fig. 1 (e) at the time of $t=10^{4}$, still like that in the
harmonic chain. The speed of the first pulse is calculated and it is a
little larger than the acoustic velocity of 1, and so it is not a linear
wave. The first pulse travels a little faster than the second pulse, so as
long as time is long enough, the first pulse will separate from the second
one. Fig. 1 (f) displays the separation of the first pulse from the others
at $t=2\times 10^{5}$, and from then on its shape and energy do not changed
any more, and it becomes a solitary wave. We show this solitary wave in the
momentum space and configuration space in Fig. 2. The velocity $v$ and
energy $E$ of the solitary wave are calculated numerically: $v=1.000743$, $%
E=0.044664$, and $v$ and $E$ satisfy the relation of Eq. (17). Then the
solitary wave may be a soliton solution. $A$ can be obtained from Eq. (12)
(Eq. (16)) with $v$ ($E$). Then we fit the solitary wave with Eq. (10) and
(9) in Fig. 2. and the excellent fit shows that the solitary wave excited by
the small kick $p=0.7$ is a soliton indeed.

We present the width of solitary waves at different energy in Fig. 3 (a).
The energy of solitary wave is determined by the kick strength and can be
calculated with Eq. (13), where the sum is computed over the lattices on
which the solitary wave is localized. At the energy of $E<0.1$ the width of
solitary wave is relatively wide and exhibits a power-law decay with energy.
At $E>0.1$ the decay of width becomes slower and the localization of the
solitary wave becomes strong as energy increases. At $E>50$ solitary wave is
always localized on the 4 lattices, the same as that in the pure anharmonic
lattice.

Fig. 3 (b) shows the relation between energy and velocity of solitary waves
in the FPU lattice. If the energy $E<0.1$, the width of solitary wave is
large and the continuum approximation is applicable, so the relation is
consistent with Eq. (17) [see the inset in Fig. 3 (b)] and the solitary wave
is a soliton. According to Eq. (17), soliton velocity is only a little
larger than the acoustic velocity because its energy is so small. If $E>0.1$%
, the solitary wave is localized on a few lattices and the long wavelength
limit is not satisfied. This leads the deviation from the fitted curve of
Eq. (17). So with the increase of energy, the nonlinear part in the
interaction potential of Eq. (2) becomes more and more obvious and soliton
turns into solitary wave. If $E>50$, velocity increases with energy as a
power-law function, asymptotic to $v=aE^{1/4}$ of Eq. (18) in the pure
anharmonic lattice. In this case, the linear part in Eq. (2) is negligible
while the nonlinear part is significant, and the property of solitary wave
becomes similar to that in the pure anharmonic lattice.

In the following we discuss solitary wave scattering dynamics. Scattering
property is an important key to determine whether it is a soliton or a
solitary wave. Soliton scattering is elastic while solitary wave scattering
is inelastic. Fig. 4 show the process of head-on collision of a pair of
solitary wave $a$ and $b$. The energy of $a$ is one half of that of $b$,
i.e., $E_{a}/E_{b}=1/2$. Before collision each solitary wave maintains shape
and energy, as shown in Fig. 4 (a), (c), (e) and (g). After collision the
two solitary waves pass through each other, denoted as $a^{\prime }$ and $%
b^{\prime }$, and the scattering behaviors are significant different in the
following four cases. In Fig. 4 (b), the two solitary waves $a$ and $b$ with 
$E_{a}=0.01$ and $E_{b}=0.02$ do not change their shapes\ and energy and it
is an elastic scattering. So $a$ and $b$ are solitons. In Fig 4 (d),
solitary wave $a$ with $E_{a}=0.05$ is unchanged, i.e., $a^{\prime }$ is the
same as $a$, while solitary wave $b$ with $E_{b}=0.1$ is scattered to
solitary wave $b^{\prime }$ with a tapered tail. The energy of solitary wave 
$b$ decreases and $E_{b}=E_{b^{\prime }}+E_{b^{\prime }t}$, here $%
E_{b^{\prime }t}$ is the energy of the tail of $b^{\prime }$. In Fig 4 (f),
both solitary wave $a$ and $b$ with $E_{a}=0.1$ and $E_{b}=0.2$ are
scattered. They are scattered to $a^{\prime }$ and $b^{\prime }$ with
tapered tails respectively. The energy of $a$ and $b$ decrease. $%
E_{a}=E_{a^{\prime }}+E_{a^{\prime }t}$ and $E_{b}=E_{b^{\prime
}}+E_{b^{\prime }t}$, $E_{a^{\prime }t}$ is the energy of the tail of $%
a^{\prime }$. In Fig 4 (h), the scattering behavior of solitary waves with $%
E_{a}=0.3$ and $E_{b}=0.6$ is quite different from that in Fig 4 (f). Energy
exchange takes place and the scattering effect is enhanced. The large
solitary wave $b$ loses energy and the small one $a$ gains energy, and extra
wave packets are excited. So the collision process varies from elastic to
inelastic and then the scattering effect becomes stronger with the increase
of the energy of solitary wave.

Next we calculate the energy scattering rate of solitary wave to describe
the scattering behavior quantitatively. The energy of a solitary wave is the
sum of kinetic energy and potential energy. In the FPU lattice the soliton
energy is small and its width is wide. Kinetic energy and potential energy
don't varies with time while soliton travels, as shown in Fig. 5 (a), where
soliton energy is 0.01. So after collision, the scattering result does not
change with time delay of the two solitons, labeled as phase $\delta $ [see
Fig. 5 (b)]. As energy increases, soliton becomes solitary wave. Solitary
wave is only localized on a few lattices. So due to the discrete nature of
the lattice, during a solitary wave is travelling on the lattice, its total
energy is unchanged while its kinetic energy and potential energy vary with
time periodically, as presented in Fig. 5 (c), in which solitary wave energy
is 0.3. The variation period T is the time that the solitary wave spends on
moving the distance of one lattice constant. Here the lattice constant is
unity, and then $T=1/v$. This leads the scattering results also vary with
phase $\delta $ periodically \cite{13,23} [see Fig. 5 (d)]. For practical
application, statistical properties are more significant than the prompt
values. Then we investigate the average energy scattering rates, $%
<E_{a}^{\prime }-E_{a}>/E_{a}$, $<E_{b}^{\prime }-E_{b}>/E_{b}$, and the
average rate of the energy "loss", $<\Delta E>/(E_{a}+E_{b})$.

Fig. 6 present the average energy scattering rates and the average rate of
the energy "loss" with the increase of energy when $E_{a}/E_{b}$ is fixed as 
$1/2$. When the solitary wave energy $E<0.1$, the scattering rate is zero
and the scattering is elastic. The scattering process is similar to that in
Fig. 4 (a) and (b). The shapes and energy of the two solitary waves are
unchanged. So the solitary wave with energy $E<0.1$ is a soliton. When $%
0.1<E<0.3$, the energy scattering rates of $a$ and $b$ are both less than 0
and the scattering is inelastic. The scattering process can be shown by Fig.
4 (e) and (f) approximately. Both of the two solitary waves are scattered
into smaller ones with tapered tails and there is no energy exchange between
them. The scattering effect is quite weak as the scattering rate is around $%
10^{-3}$. Soliton is transformed into solitary wave with the increase of
energy. When $E>0.3$, the energy scattering rate is more than 0 for the
small solitary wave $a$ and is less than 0 for the large solitary wave $b$,
and the scattering is inelastic. The scattering process is like that in Fig.
4 (g) and (h). There is energy exchange between the two solitary waves,
i.e., large solitary wave $b$ loses energy and small one $a$ obtains energy
on average, and extra wave packets are excited. The stars in Fig. 6 are the
scattering results for the pure anharmonic lattice. Any pair of solitary
waves with the same ratio $E_{a}/E_{b}$ has the same scattering rates due to
the scaling property \cite{13}. In the FPU lattice the scattering effect
increases with energy and the scattering results approach those in the pure
anharmonic lattice when $E\rightarrow \infty $.

\textbf{V.\bigskip\ CONCLUSION AND DISCUSSION}

In summary, we explore the transition from solitons to solitary waves in the
FPU lattice. Solitons and solitary waves can be excited by momentum kicks.
If energy is smaller than 0.1, the excitation is a soliton, which is
consistent with analytical results. Solitons maintain shape and energy, and
kinetic and potential energy remain unchanged while travelling. The soliton
scattering is elastic; solitons do not alter shape and energy after
collision with others. If energy is larger than the threshold of 0.1,
soliton transforms to solitary wave. When a solitary wave propagates,
kinetic energy and potential energy are varying periodically, while total
energy does not change. The solitary wave scattering is inelastic; both of
solitary waves lose energy, or large solitary wave loses energy and small
one gains on average, and extra wave packets are excited. If energy is large
enough, the properties of solitary waves tend to those in the pure
anharmonic lattice.

The heat conduction behaviors of the FPU lattice are quite different at low
and high temperature. The temperature gradient can't be formed like the
harmonic lattice at low temperature, while it can be set up at high
temperature \cite{12,24,25}. We discuss it with the concept of solitons and
solitary waves. In molecular dynamics simulations, the temperature at
position $n$ is defined as $T_{n}=$ $<p_{n}^{2}>/2$ and the role of a
thermostat can be interpreted as a series of kicks on the ends of the
lattice. When the thermostat temperature is low, solitons are excited. The
soliton scattering is elastic and there is no energy exchange, and so there
is no temperature gradient formed. In the case of high thermostat
temperature, solitary waves are obtained. The solitary wave scattering is
inelastic and energy exchange takes place, and so temperature gradient is
set up.

\bigskip \textbf{ACKNOWLEDGMENT}

This work is supported by the National Natural Science Foundation of China
under grant No. 10805034.

\begin{equation*}
\text{Figure Captions}
\end{equation*}

FIG. 1 The momentum $p_{i}$ versus lattice position $i$. (a) the harmonic
chain at $p_{1}=0.7$ and $t=500$. (b) the pure anharmonic chain at $%
p_{1}=0.7 $ and $t=500$. (c) the FPU chain at $p_{1}=1.5$ and $t=500$. (d),
(e) and (f) represent the FPU chain at $p_{1}=0.7$ with $t=500$, $t=10^{4}$
and $t=2\times 10^{5}$, respectively.

FIG. 2 Dots represent the solitary wave in Fig. 1 (f) in the momentum space
(a) and configuration space (b). Solid lines in (a) and (b) stand for fitted
curves of Eq. (10) and Eq. (9), respectively.

FIG. 3 The width (a) and velocity (b) versus energy of solitary waves in
log-log scale (dots). (a) $x$ axis is the energy of solitary wave $E$ and $y$
axis is the number $N$ of lattices where solitary wave is localized. (b) The
solid line is a fit of Eq. (17), and the dashed line is a fit of Eq. (18).
The inset enlarges the region indicated by the dotted-line rectangle.

FIG. 4 Head-on collision processes of a pair of solitary wave $a$ and $b$.
(a) and (b), $E_{a}=0.01$ and $E_{b}=0.02$. (c) and (d), $E_{a}=0.05$ and $%
E_{b}=0.1$. (e) and (f), $E_{a}=0.1$ and $E_{b}=0.2$. (g) and (h), $%
E_{a}=0.3 $ and $E_{b}=0.6$. (a), (c), (e) and (g) are before collision and
(b), (d), (f) and (h) are after collision. Insets in (d), (f) and (h) are
enlargement of rectangle.

FIG. 5 (a) and (c), kinetic energy (solid line) and potential energy (dot
line) versus time. Soliton energy is 0.01 in (a) and solitary wave energy is
0.3 in (c). (b) and (d), the energy of scattered solitary wave $a^{\prime }$
versus phase $\delta $. (b) is corresponding to the collision process in
Fig. 4 (a) and (b), and (d) is for the process in Fig. 4 (g) and (h).

FIG. 6 The average energy scattering rates and the average rate of the
energy "loss" at $E_{a}/E_{b}=1/2$ on a semilogarithmic scale. (a), $%
<E_{a}^{\prime }-E_{a}>/E_{a}$ versus $E_{a}$. Inset is enlargement of
rectangle. (b), $<E_{b}^{\prime }-E_{b}>/E_{b}$ versus $E_{b}$. (c), $%
<\Delta E>/(E_{a}+E_{b})$ versus $E_{a}+E_{b}$. The dot and star represent
the results of solitary waves in the FPU lattice and in the pure anharmonic
lattice, respectively.

\bigskip

*Electronic address: wenzy@scu.edu.cn


\bigskip

\begin{thebibliography}{99}
\bibitem{1} J. S. Russell, Report on Waves, Report of the Meeting of the
British Association for the Advancement of Science (John Murray, London,
1844).

\bibitem{2} D. J. Korteweg and G. de Vries, Philos. Mag. \textbf{539}, 422
(1895).

\bibitem{3} E. Fermi, J. Pasta, and S. Ulam, Los Alamos Document No. LA-1940
(1955).

\bibitem{4} N. J. Zabusky and M. D. Kruskal, Phys. Rev. Lett. \textbf{15},
240 (1965).

\bibitem{5} A. C. Scott, \textit{Nonlinear Science} (Oxford University
Press, New York, 1999).

\bibitem{6} \textit{The Encyclopedia of Nonlinear Science}, edited by A. C.
Scott (Routledge, New York, 2005).

\bibitem{7} M. Remoissenet, \textit{Waves Called Solitons} (Springer,
Berlin, 1999).

\bibitem{8} G. Friesecke and R. L. Pego, Nonlinearity \textbf{12}, 1601
(1999).

\bibitem{9} G. Friesecke and R. L. Pego, Nonlinearity \textbf{15}, 1343
(2002).

\bibitem{10} G. Friesecke and R. L. Pego, Nonlinearity \textbf{17}, 207
(2004).

\bibitem{11} G. Friesecke and R. L. Pego, Nonlinearity \textbf{17}, 229
(2004).

\bibitem{12} B. Hu, B. Li, and H. Zhao, Phys. Rev. E \textbf{61}, 3828
(2000).

\bibitem{13} H. Zhao, Z. Wen, Y. Zhang, and D. Zheng, Phys. Rev. Lett. 
\textbf{94}, 025507 (2005).

\bibitem{14} R. Khomeriki, Phys. Rev. E \textbf{65}, 026605 (2002)

\bibitem{15} A. Rosas and K. Lindenberg, Phys. Rev. E \textbf{69, }016615
(2004).

\bibitem{16} G. S. Zavt, M. Wagner, and A. L$\overset{..}{u}$tze, Phys. Rev.
E \textbf{47}, 4108 (1993).

\bibitem{17} Z. Yuan, J. Wang, M. Chu, G. Xia, and Z Zheng, Phys. Rev. E 
\textbf{88}, 042901 (2013).

\bibitem{18} Y. Kosevich, Phys. Rev. B \textbf{47}, 3138 (1993).

\bibitem{19} P. Poggi, S. Ruffo and H. Kantz, Phys. Rev. E \textbf{52}, 307
(1995)

\bibitem{20} M.J. Ablowitz and H. Segur, \textit{Solitons and the Inverse
Scattering Transform} (SIAM, Philadelphia, 1981).

\bibitem{21} M.J. Ablowitz and P.A. Clarkson, \textit{Solitons, Nonlinear
Evolution Equations and Inverse Scattering} (Cambridge University Press,
Cambridge, 1991)

\bibitem{22} J. Szeftel, Pascal Laurent-Gengoux, and E. Ilisca, Phys. Rev.
Lett. \textbf{83}, 3982 (1999).

\bibitem{23} Z. Wen and H. Zhao, Chin. Phys. Lett. \textbf{22}, 1340 (2005).

\bibitem{24} B Li, H Zhao, and B Hu, Phys. Rev. Lett. 86, 63 (2001)

\bibitem{25} S. Lepri, R. Livi, and A. Politi, Phys. Rev. Lett. 78, 1896
(1997)
\end{thebibliography}
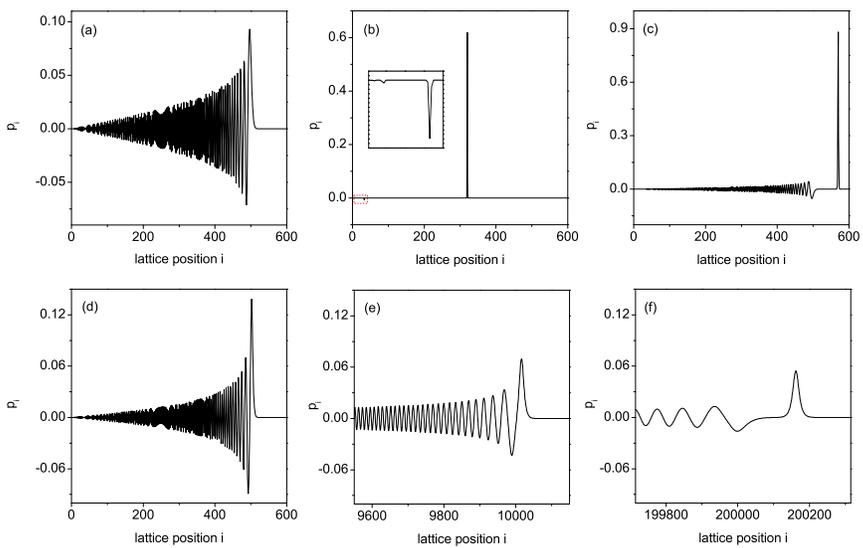
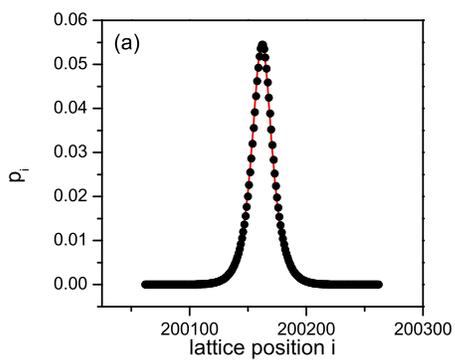
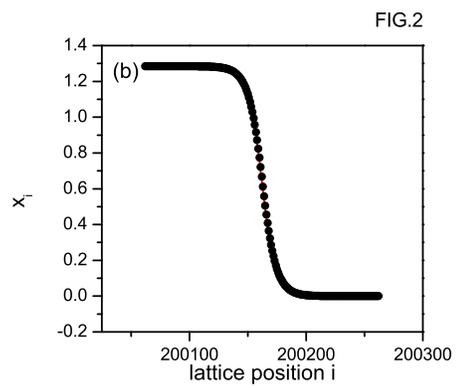
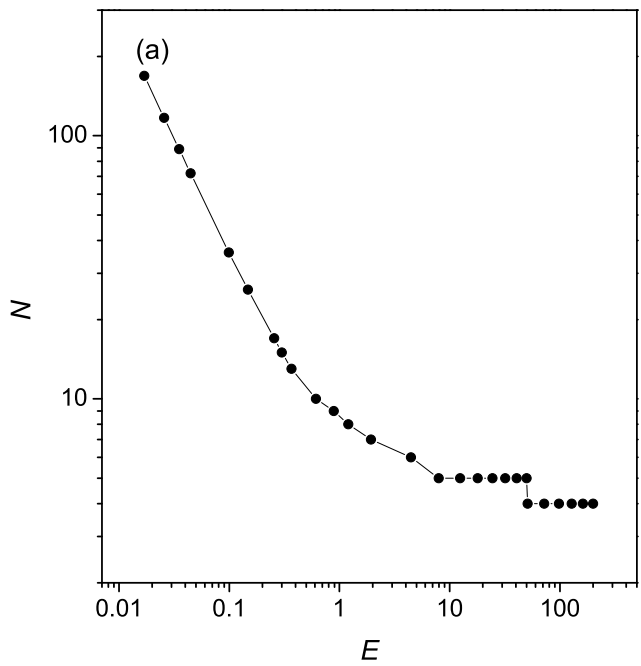
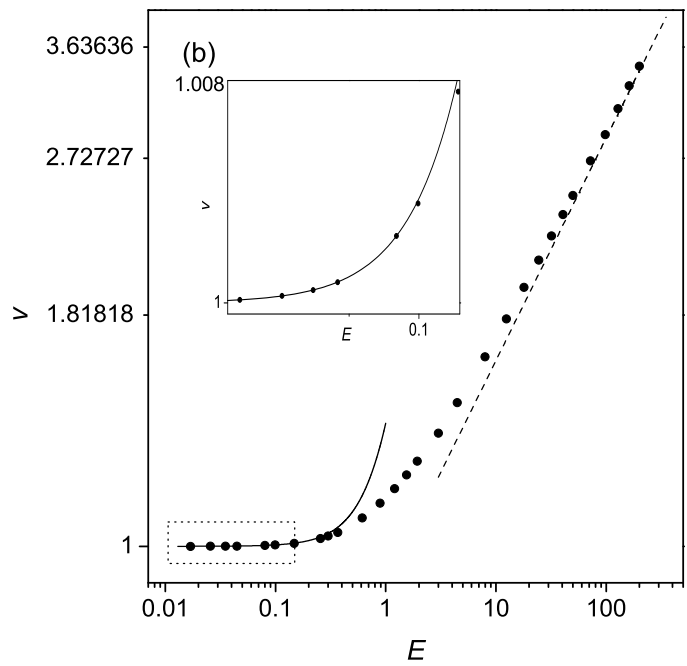
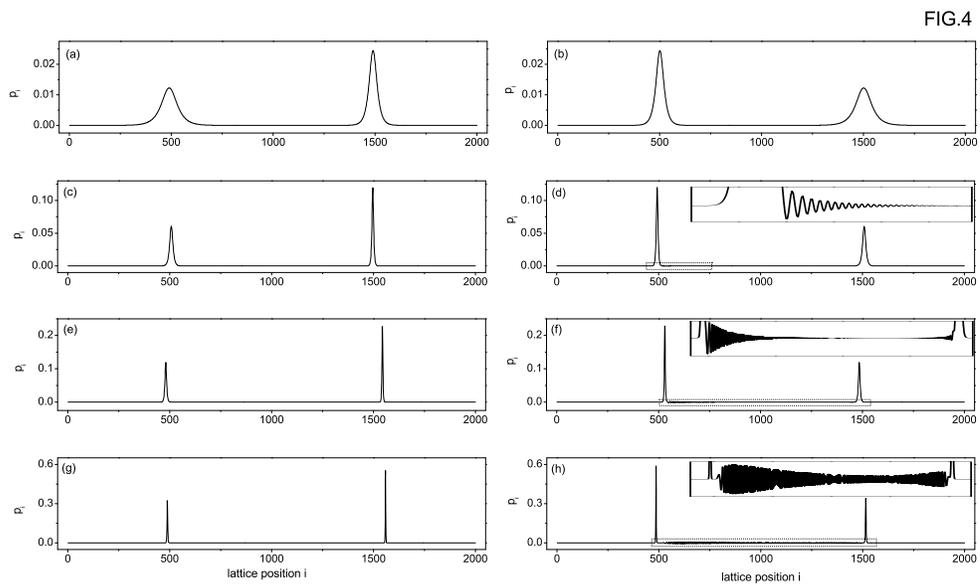
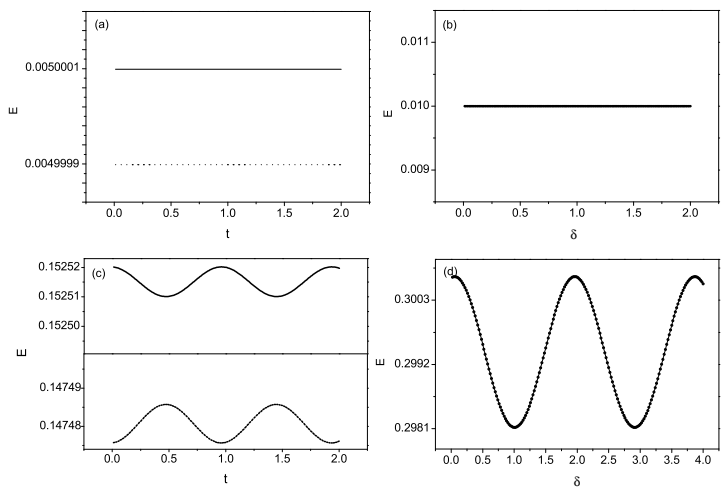
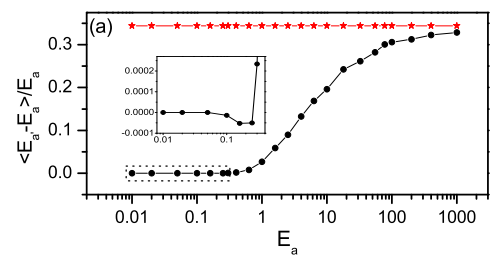
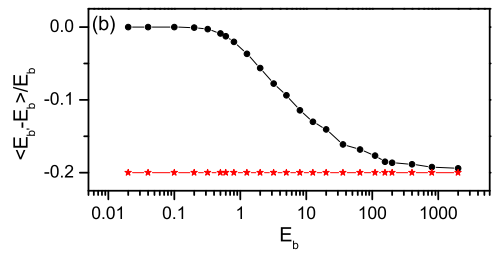
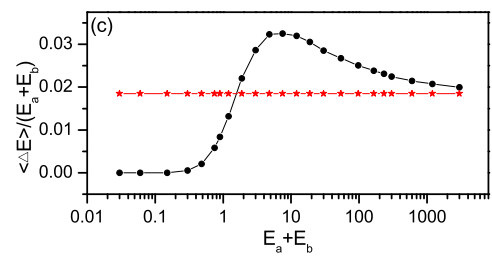
\end{document}